\documentclass[letterpaper,twocolumn,prl,aps,unsortedaddress]{revtex4-1}

\usepackage[T1]{fontenc}
\usepackage[latin9]{inputenc}
\usepackage{color}
\usepackage{float}
\usepackage{amsmath}
\usepackage{amssymb}
\usepackage{graphicx}

\makeatletter

\@ifundefined{textcolor}{}
{%
 \definecolor{BLACK}{gray}{0}
 \definecolor{WHITE}{gray}{1}
 \definecolor{RED}{rgb}{1,0,0}
 \definecolor{GREEN}{rgb}{0,1,0}
 \definecolor{BLUE}{rgb}{0,0,1}
 \definecolor{CYAN}{cmyk}{1,0,0,0}
 \definecolor{MAGENTA}{cmyk}{0,1,0,0}
 \definecolor{YELLOW}{cmyk}{0,0,1,0}
}

\@ifundefined{definecolor}{\@ifundefined{definecolor}
 {\usepackage{color}}{}
}{}%
\usepackage{graphics}\usepackage{bm}\usepackage{psfrag}\usepackage{epsfig}\usepackage{grffile}\usepackage{subfigure}
\newcommand{\beq}{\begin{equation}}\newcommand{\enq}{\end{equation}}\newcommand{\ceq}[1]{(\ref{#1})}\newcommand{\kk}{{\bf k}}

\makeatother

\begin{document}

\title{Chiral superfluid states in hybrid graphene heterostructures}

\author{Junhua Zhang and E. Rossi}

\affiliation{ Department of Physics, College of William and Mary, Williamsburg,
VA 23187, USA }

\date{\today}
\begin{abstract}
We study the {}``hybrid''
heterostructure formed by one sheet of single layer graphene (SLG)
and one sheet of bilayer graphene (BLG) separated by a thin film of
dielectric material. In general it is expected that interlayer interactions
can drive the system to a spontaneously broken symmetry state characterized
by interlayer phase coherence. The peculiarity of the SLG-BLG heterostructure
is that the electrons in the two layers have different chiralities.
We find that this difference
causes the spontaneously broken symmetry state to be N-fold
degenerate. Moreover, we find that some of the degenerate states are
chiral superfluid states, topologically distinct from the usual layer-ferromagnetism.
The chiral nature of the ground state opens the possibility to realize
protected midgap states. The N-fold degeneracy of the ground state
makes the physics of SLG-BLG hybrid systems analogous to the physics
of $^{3}{\rm He}$, in particular given the recent discovery of chiral
superfluid states in this system \cite{pollanen2012}. 
\end{abstract}

\maketitle

Graphene \cite{novoselov2004} and bilayer graphene \cite{novoselov2006}
are ideal 2D electronic systems 
\cite{neto2009,dassarma2011}
in which the conduction and valence
bands touch at single points, charge neutrality points, at
the corners of the Brillouin zone (BZ). Around these points the low
energy electronic states are well described as massless Dirac fermions
with Berry phase $\pi$ in SLG and as massive chiral fermions with
Berry phase $2\pi$ in BLG. 
Recently, the use of hexagonal boron nitride (hBN) films \cite{dean2010}
has allowed  the realization of
graphene 
heterostructures \cite{britnell2012,haigh2012}
in which the graphene layers are only few nanometers apart and still electrically
isolated \cite{kim2009,kim2010,ponomarenko2011,kim2012,gorbachev2012}.
In this situation interlayer interactions 
can drive the system into an interlayer phase coherent ground
state \cite{min2008,zhang2008jog,lozovik2008,kharitonov2008b}. This
state can be thought of as an exciton condensate \cite{keldysh1965,lozovik1975}
of electrons in one layer and holes in the other layer, as a superfluid
state \cite{lutchyn2010}, or by treating the layer degree of freedom as a spin degree
of freedom ({\em pseudospin}) as a ferromagnetic state. 
Experimental evidence suggests that 
the interlayer phase coherent state has been realized 
in quantum Hall bilayers
\cite{eisenstein2004,spielman2000,spielman2001,kellogg2002,kellogg2003,champagne2008,tutuc2004}
and very recently \cite{gorbachev2012} 
in symmetric
double layer graphene systems.
The experimental capability to realize high quality graphene-hBN heterostructures
has made possible to study the effects of interactions between fermionic quasiparticles having
qualitatively different dispersion and chirality. This can be
realized by creating heterostructures in which one layer is SLG
and the other BLG.

In this Letter, we study the nature of the interlayer broken symmetry
state for SLG-hBN-BLG systems. We find that the difference in
the dispersion and chirality between the two layers profoundly modifies
the nature of the ground state. In particular, we find that due to
the difference of chirality: (\textbf{i}) the interlayer broken symmetry
state is N-fold degenerate (N=2 or 4 depending on the nature, long-range or short-range,
of the interlayer interaction); (\textbf{ii}) one of the
degenerate states is always chiral, i.e characterized by a complex
order parameter whose phase depends on the momentum direction. 
The N-fold degeneracy
of the ground state raises the possibility that in SLG-BLG systems a 
state could be realized analogous to states realized in $^{3}{\rm He}$
\cite{leggett1975}. Moreover, the chiral nature of one of the ground
states makes possible the realization of protected midgap states in
the presence of vortices in the exciton condensate \cite{balatsky2004,seradjeh2008,seradjeh2009}.

\begin{figure}[htb]
\includegraphics[width=8.11cm]{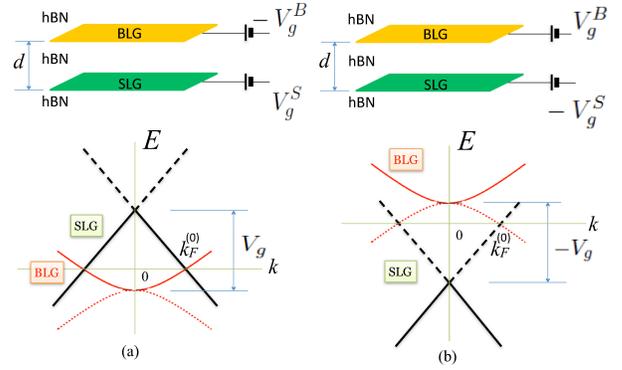}

\caption{(Color online). (a),
SLG and BLG are gated individually at voltages $V_{g}^{S}$ and $-V_{g}^{B}$, $V_{g}=V_{g}^{S}+V_{g}^{B}$.
At low energies and low voltages the most relevant bands are the BLG conduction
band and the SLG valence band.
(b), By inverting the voltages ($-V_{g}^{S}$ and $V_{g}^{B}$)
the most relevant bands become the SLG conduction band and
the BLG valence band. \label{fig:setup} }
\end{figure}

The heterostructure that we study is shown schematically in Fig.~\ref{fig:setup}.
The two layers are connected to separate gates
($V_{g}^{S},-V_{g}^{B}$) so that their doping can be controlled
independently and can be adjusted to have the p-type Fermi surface
(FS) in one layer nested with the n-type FS in the other, condition
that favors the instability toward the formation of the exciton condensate.
Let $V_{g}=V_{g}^{B}+V_{g}^{S}$
be the bias voltage for which the FS's in BLG and SLG are nested.
At low energies the band structure of
SLG is well described by two inequivalent valleys (at the $K$ and
$K'$ points in the BZ) around which the fermionic dispersion is linear.
In BLG the low energy conduction and valence bands also touch at the
points $K$ $K'$ but around these points the dispersion is nearly
parabolic with an effective mass $m\approx0.03\, m_{e}$ \cite{neto2009,dassarma2011}.
For this experimental setup the effective low-energy band structure is formed
by the conduction band of BLG and the valence band of SLG (or vice
versa as shown in Fig.~\ref{fig:setup}~(b)).

The low energy physics of the SLG-BLG system is described by the Hamiltonian:
$\mathcal{H}=\mathcal{H}_{0}+\mathcal{H}_{int},$ where,
in the limit of vanishing interlayer tunneling,
the noninteracting
Hamiltonian $\mathcal{H}_{0}=\sum_{\mathbf{k},\sigma}\varepsilon_{\mathbf{k},\sigma}c_{\mathbf{k},\sigma}^{\dagger}c_{\mathbf{k},\sigma}$
with $\sigma=1,2$ representing the layer degree of freedom treated
as a pseudospin. $c_{\mathbf{k},\sigma}^{\dagger}$ ($c_{\mathbf{k},\sigma}$)
is the creation (annihilation) operator for a fermion with momentum
$\mathbf{k}$ in layer $\sigma$. Assuming, for concreteness, that
the gate voltages are such that the Fermi energy lies in the conduction
band for BLG ($\sigma=1)$ and in the valence band for SLG ($\sigma=2$)
we have $\varepsilon_{\mathbf{k},1}=-V_{g}^{B}+[\hbar^{2}v_{F}^{2}k^{2}+\frac{\gamma_{1}^{2}}{2}-[\frac{\gamma_{1}^{4}}{4}+\hbar^{2}v_{F}^{2}k^{2}\gamma_{1}^{2}]^{1/2}]^{1/2}$
and $\varepsilon_{\mathbf{k},2}=V_{g}^{S}-\hbar v_{F}k$ to which
correspond the eigenstates $\psi_{\mathbf{k},1}=\frac{1}{\sqrt{2}}\left(1,e^{i\eta m\theta_{\mathbf{k}}}\right)^{T}$
and $\psi_{\mathbf{k},2}=\frac{1}{\sqrt{2}}\left(1,-e^{i\eta n\theta_{\mathbf{k}}}\right)^{T}$
respectively, where $m=2$ for BLG and $n=1$ for SLG are the integers
that specify the chirality of the two layers, and $\eta=+1\ (-1)$
for states around the $K$ ($K'$) point.
Below we consider the states
around the $K$ point only as the $K'$ point follows similar analysis.
$v_{F}\approx10^{6}$~m/s is the Fermi velocity of SLG close to the
Dirac point, $\gamma_{1}\approx400\,\mathrm{meV}$, and
$\theta_{\mathbf{k}}\equiv\arctan(k_{y}/k_{x})$.
The form of $\mathcal{H}_{0}$ that we use is valid as long
as  $|V_g^S|<140$~meV and $3$~meV$\lesssim |V_g^B|\lesssim 200$~meV
\cite{mccann2006,dassarma2011}.
For the interacting part of $\mathcal{H}$ we have: 
\begin{align}
\mathcal{H}_{int} & =\frac{1}{2A}\sum_{\sigma}\sum_{\mathbf{k,k',q}}V_{\mathbf{q}}f_{\sigma}(\theta_{\mathbf{k+q}}-\theta_{\mathbf{k}})f_{\sigma}(\theta_{\mathbf{k'-q}}-\theta_{\mathbf{k'}})\times\nonumber \\
 & \ \ \ \ \ \ \ \ \ \ c_{\mathbf{k+q},\sigma}^{\dagger}c_{\mathbf{k'-q},\sigma}^{\dagger}c_{\mathbf{k'},\sigma}c_{\mathbf{k},\sigma}+\nonumber \\
 & +\frac{1}{A}\sum_{\mathbf{k,k',q}}V_{\mathbf{q}}^{d}f_{1}(\theta_{\mathbf{k+q}}-\theta_{\mathbf{k}})f_{2}(\theta_{\mathbf{k'-q}}-\theta_{\mathbf{k'}})\times\nonumber \\
 & \ \ \ \ \ \ \ \ \ \ c_{\mathbf{k+q},1}^{\dagger}c_{\mathbf{k'-q},2}^{\dagger}c_{\mathbf{k'},2}c_{\mathbf{k},1},\label{eq:miscroscopicHamiltonian}
\end{align}
where $A$ denotes the area of the heterostructure, $V_{\mathbf{q}}^{d}$
($V_{\mathbf{q}}$) refers to the interlayer (intralayer) interaction,
and $f_{1}(\theta_{\mathbf{k}}-\theta_{\mathbf{p}})=\frac{1}{2}\left[1+e^{2i(\theta_{\mathbf{k}}-\theta_{\mathbf{p}})}\right],\ f_{2}(\theta_{\mathbf{k}}-\theta_{\mathbf{p}})=\frac{1}{2}\left[1+e^{i(\theta_{\mathbf{k}}-\theta_{\mathbf{p}})}\right]$
are factors that arise from the wavefunction overlap between states
$\psi_{\mathbf{k},\sigma}$, $\psi_{\mathbf{p},\sigma'}$.

To decouple the interactions we use the Hartree-Fock approximation
and obtain the mean-field Hamiltonian 
\begin{equation}
\mathcal{H}_{MF}=\sum_{\mathbf{k},\sigma,\sigma'}c_{\mathbf{k},\sigma}^{\dagger}\left(\Delta_{\mathbf{k}}^{0}\tau_{\sigma\sigma'}^{0}-\boldsymbol{\Delta}_{\mathbf{k}}\cdot\boldsymbol{\tau}_{\sigma\sigma'}\right)c_{\mathbf{k},\sigma'},\label{eq: mean-field Hamiltonian}
\end{equation}
where $[\Delta^{0},\boldsymbol{\Delta}=(\Delta^{x},\Delta^{y},\Delta^{z})]$
are the mean-fields and $[\tau^{0},\boldsymbol{\tau}=(\tau^{x},\tau^{y},\tau^{z})]$
the $2\times2$ identity and Pauli matrices acting in the layer
pseudospin space. Due to the asymmetry of the band dispersion between
the two layers the field $\Delta_{\mathbf{k}}^{0}$ does not vanish,
unlike in symmetric double-layer systems. The transverse components
of the pseudospin field $\boldsymbol{\Delta}_{\mathbf{k}}$
form a complex order parameter $\Delta_{\mathbf{k}}^{\perp}=\Delta_{\mathbf{k}}^{x}-i\Delta_{\mathbf{k}}^{y}$,
whose magnitude $\left|\Delta_{\mathbf{k}}^{\perp}\right|$ measures
the strength of the particle-hole condensate. The mean-fields 
are given by the following self-consistent
equations: 
\begin{align}
 & \Delta_{\mathbf{k}}^{0}=(\varepsilon_{\mathbf{k},2}+\varepsilon_{\mathbf{k},1})/2+\nonumber \\
 & +\frac{1}{2A}\sum_{\mathbf{p}}\left[V_{\mathbf{k-p}}F_{1}(\theta_{\mathbf{k-p}})+\frac{2\pi e^{2}}{\epsilon}gd\right]\left(1-n_{\mathbf{p}}^{-}-n_{\mathbf{p}}^{+}\right)\nonumber \\
 & -\frac{1}{2A}\sum_{\mathbf{p}}V_{\mathbf{k-p}}F_{2}(\theta_{\mathbf{k-p}})\left[1+\frac{\Delta_{\mathbf{p}}^{z}}{E_{\mathbf{p}}}\left(n_{\mathbf{p}}^{-}-n_{\mathbf{p}}^{+}\right)\right];\label{eq:D0}
\end{align}
\begin{align}
 & \Delta_{\mathbf{k}}^{z}=(\varepsilon_{\mathbf{k},2}-\varepsilon_{\mathbf{k},1})/2+\nonumber \\
 & \frac{1}{2A}\sum_{\mathbf{p}}\left[V_{\mathbf{k-p}}F_{1}(\theta_{\mathbf{k-p}})-\frac{2\pi e^{2}}{\epsilon}gd\right]\left[1+\frac{\Delta_{\mathbf{p}}^{z}}{E_{\mathbf{p}}}\left(n_{\mathbf{p}}^{-}-n_{\mathbf{p}}^{+}\right)\right]\nonumber \\
 & -\frac{1}{2A}\sum_{\mathbf{p}}V_{\mathbf{k-p}}F_{2}(\theta_{\mathbf{k-p}})\left(1-n_{\mathbf{p}}^{-}-n_{\mathbf{p}}^{+}\right);\label{eq:Dz}
\end{align}
\begin{equation}
\Delta_{\mathbf{k}}^{\perp}=\frac{1}{2A}\sum_{\mathbf{p}}V_{\mathbf{k-p}}^{d}F^{d}(\theta_{\mathbf{k-p}})\left[\frac{\Delta_{\mathbf{p}}^{\perp}}{E_{\mathbf{p}}}\left(n_{\mathbf{p}}^{-}-n_{\mathbf{p}}^{+}\right)\right];\label{eq:Dperp}
\end{equation}
where $g=4$ is the total spin and valley degeneracy and $\epsilon$
the dielectric constant of the embedding media. The $\frac{2\pi e^{2}}{\epsilon}gd$
term is specific to the interlayer Coulomb interaction in the direct
channel. $n_{\mathbf{p}}^{\pm}=1/\left[\exp\left(\varepsilon_{\mathbf{p}}^{\pm}/k_{B}T\right)+1\right]$
are the occupation numbers at temperature $T$ of the renormalized
bands with band energies $\varepsilon_{\mathbf{k}}^{\pm}=\Delta_{\mathbf{k}}^{0}\pm E_{\mathbf{k}}$,
where $E_{\mathbf{k}}=\left[\left(\Delta_{\mathbf{k}}^{z}\right)^{2}+\left|\Delta_{\mathbf{k}}^{\perp}\right|^{2}\right]^{1/2}$,
and $F_{1}(\theta_{\mathbf{k-p}})$, $F_{2}(\theta_{\mathbf{k-p}})$,
$F^{d}(\theta_{\mathbf{k-p}})$ (with $\theta_{\mathbf{k-p}}\equiv\theta_{\mathbf{k}}-\theta_{\mathbf{p}}$)
are angle-dependent {\em chiral-factors}. Specifically, the intralayer
chiral factors have the expressions $F_{1}(\theta_{\mathbf{k-p}})=\frac{1}{4}\left(\cos2\theta_{\mathbf{k-p}}+\cos\theta_{\mathbf{k-p}}+2\right)$
and $F_{2}(\theta_{\mathbf{k-p}})=\frac{1}{4}\left(\cos2\theta_{\mathbf{k-p}}-\cos\theta_{\mathbf{k-p}}\right)$
for SLG-BLG, whereas the interlayer chiral factor can be written in
a general form as 
\begin{equation}
F^{d}\left(\theta_{\mathbf{k-p}}\right)=\frac{1}{4}\left(e^{-in\theta_{\mathbf{k-p}}}+e^{0}+e^{i(m-n)\theta_{\mathbf{k-p}}}+e^{im\theta_{\mathbf{k-p}}}\right).\label{eq: generic_chiral}
\end{equation}
In the SLG-SLG structure $m=n=1$, in the hybrid SLG-BLG
structure $m\neq n$ with $m=2$ and $n=1$.

To understand the consequence of the difference in the chiral-factor
$F^{d}(\theta_{\mathbf{k-p}})$ on the gap equation between the symmetric
SLG-SLG heterostructure and the asymmetric SLG-BLG, let us write the
general solution of the gap equation \ceq{eq:Dperp} as $\Delta_{\mathbf{k}}^{\perp}=\left|\Delta_{\mathbf{k}}^{\perp}\right|_{J}e^{iJ\theta_{\mathbf{k}}+i\phi}$
with the chirality $J=0,\pm1,\pm2,\dots$ and an arbitrary global
phase $\phi$. Without loss of generality we assume $\Delta_{\mathbf{k}}^{0}$,
$\Delta_{\mathbf{k}}^{z}$, and the magnitude $\left|\Delta_{\mathbf{k}}^{\perp}\right|$
to be angle-independent (it is straightforward to verify that this
assumption is consistent with the self-consistent mean-field equations).
The gap equation \ceq{eq:Dperp} becomes 
\begin{align}
 & \left|\Delta_{\mathbf{k}}^{\perp}\right|_{J}\nonumber \\
 & =\frac{1}{2A}\sum_{\mathbf{p}}V_{\mathbf{k-p}}^{d}F^{d}(\theta_{\mathbf{k-p}})e^{-iJ\theta_{\mathbf{k-p}}}\left[\frac{\left|\Delta_{\mathbf{p}}^{\perp}\right|_{J}}{E_{\mathbf{p}}}\left(n_{\mathbf{p}}^{-}-n_{\mathbf{p}}^{+}\right)\right].\label{eq:gap02}
\end{align}
In the case of short-range interactions, $V_{\mathbf{k-p}}^{d}=\text{const}$,
from \ceq{eq:gap02} we have that in symmetric systems, such
as SLG-SLG, in which $m=n$, for $J=0$, because of
the form of the chiral factor, the effective interaction is
twice stronger than for $J\neq0$ and therefore that the
non-chiral $J=0$ state has a critical temperature higher than that of chiral  $J\neq0$
states. On the contrary, for asymmetric systems in which $m\neq n$,
such as SLG-BLG, the chiral factor (\ref{eq: generic_chiral}) ensures
that the effective interaction is the same for all the 4 states $J=-n,0,m-n,m$.
As a consequence, for heterostructures like SLG-BLG in which $m\neq n$,
in the presence of short-range interactions, the $J=-n,0,m-n,m$ states
satisfy the same gap equation and therefore, at the mean-field level,
the interlayer phase coherent ground state is 4-fold degenerate.

In many cases of interest we expect that the interactions are not
short-range but still ``central'', i.e. depending only on the magnitude
$\left|\mathbf{k-p}\right|$. In this case the parts on the right
hand side of equation~\ceq{eq:gap02} that are odd in $\theta_{\mathbf{k-p}}$
vanish after integrating over the angle and the gap equation takes
the form: 
\begin{align}
 & \left|\Delta_{\mathbf{k}}^{\perp}\right|_{J}=\frac{1}{2A}\sum_{\mathbf{p}}V_{\mathbf{k-p}}^{d}\left[\frac{\left|\Delta_{\mathbf{p}}^{\perp}\right|_{J}}{E_{\mathbf{p}}}\left(n_{\mathbf{p}}^{-}-n_{\mathbf{p}}^{+}\right)\right]\times\nonumber \\
 & \times\frac{1}{4}\biggl[\cos\Bigl((n+J)\theta_{\mathbf{k-p}}\Bigr)+\cos\Bigl(J\theta_{\mathbf{k-p}}\Bigr)+\nonumber \\
 & +\cos\Bigl((m-n-J)\theta_{\mathbf{k-p}}\Bigr)+\cos\Bigl((m-J)\theta_{\mathbf{k-p}}\Bigr)\biggl].\label{eq:gap03}
\end{align}
Equation \ceq{eq:gap03} shows that for symmetric heterostructures,
i.e., $m=n$, in the case of ``central'' interactions the $J=0$
state again has the highest effective pairing strength and therefore
the highest critical temperature \cite{min2008}. On the other hand
for asymmetric heterostructures in which $m=2n$ the states $J=0$
and $J=n$ ($J=0$ and $J=-n$ for the other valley) have the same
and the strongest pairing strength and therefore the ground state
is 2-fold degenerate.
Similarly, we find that the free energy is the same for each of the
degenerate states

\begin{figure}[htb]
\includegraphics[width=8.11cm]{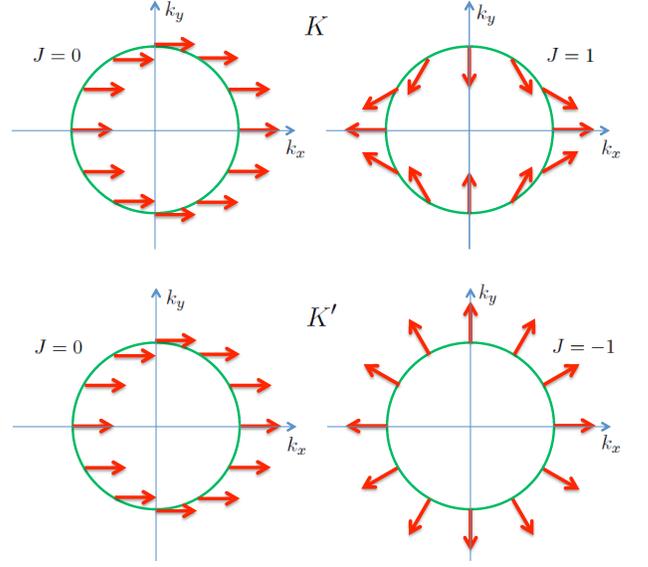}

\caption{(Color online). Pseudospin configuration
on the Fermi surface in the broken-symmetry state for a hybrid SLG-BLG
graphene heterostructure around the $K$-point (top) and the $K'$-point
(bottom). Here we have chosen $\phi=0$.
Top: the $J=0$ state: $(\Delta_{\mathbf{k}}^{x},\Delta_{\mathbf{k}}^{y})=\left|\Delta_{\mathbf{k}}^{\perp}\right|(1,0)$,
and the chiral $J=1$ state: $(\Delta_{\mathbf{k}}^{x},\Delta_{\mathbf{k}}^{y})=\left|\Delta_{\mathbf{k}}^{\perp}\right|(\cos\theta_{\mathbf{k}},-\sin\theta_{\mathbf{k}})$,
are degenerate around the $K$-point. Bottom: the $J=0$ state and
the chiral $J=-1$ state: $(\Delta_{\mathbf{k}}^{x},\Delta_{\mathbf{k}}^{y})=\left|\Delta_{\mathbf{k}}^{\perp}\right|(\cos\theta_{\mathbf{k}},+\sin\theta_{\mathbf{k}})$
are degenerate around the $K'$-point. \label{fig:c-states} }
\end{figure}

For the SLG-BLG heterostructure, in the presence of
Coulomb interactions, $V_{\mathbf{k-p}}^{d}=\frac{2\pi e^{2}}{\epsilon}\frac{e^{-\left|\mathbf{k-p}\right|d}}{\left|\mathbf{k-p}\right|}$,
we therefore find that the ground state is two-fold degenerate: around
the $K$ ($K'$) point, the non-chiral $J=0$ interlayer phase coherent
state (layer-ferromagnetic state) is degenerate with the chiral $J=1$
($J=-1$) state, see  Fig.~\ref{fig:c-states}. 
By inverting the gate voltage $V_{g}$ 
the values of $J$ at the $K$
and $K'$ points are interchanged. We find that the nature, chiral
or not chiral, of the ground state strongly affects the dynamical
density-density response function for frequencies $\omega\approx2|\Delta_\kk^\perp|$
\cite{sm} and therefore that optical measurements should
be able to distinguish between the two degenerate states.

We emphasize that the degeneracy and chirality of the phase coherent states
are due to presence of the chiral factor $F^{d}$ in the gap equation (Eq.~(\ref{eq:Dperp}))
and do not depend on the details of the band-structures of the two layers.

The fact that one of the possible interlayer phase coherent states
is chiral opens the possibility to create topologically protected
midgap states \cite{read2000,seradjeh2008,seradjeh2009} at the center
of vortices that can be created in the exciton condensate via the
{\em axial gauge field}
\cite{balatsky2004}. To see
this we observe that we can separate the mean-field Hamiltonian into
two parts $\mathcal{H}_{MF}=\mathcal{H}_{1}+\mathcal{H}_{2}$ with
$\mathcal{H}_{1}=\sum_{\mathbf{k},\sigma,\sigma'}c_{\mathbf{k},\sigma}^{\dagger}\left(\Delta_{\mathbf{k}}^{0}\tau_{\sigma\sigma'}^{0}\right)c_{\mathbf{k},\sigma'},\ \mathcal{H}_{2}=-\sum_{\mathbf{k},\sigma,\sigma'}c_{\mathbf{k},\sigma}^{\dagger}\left(\boldsymbol{\Delta}_{\mathbf{k}}\cdot\boldsymbol{\tau}_{\sigma\sigma'}\right)c_{\mathbf{k},\sigma'}$.
Since $\mathcal{H}_{1}$ and $\mathcal{H}_{2}$ commute, the eigenvalues
of $\mathcal{H}_{MF}$ are given by the sum of the eigenvalues of
$\mathcal{H}_{1}$ and $\mathcal{H}_{2}$. $\mathcal{H}_{2}$ has
a symmetric spectrum $\{\pm E_{\mathbf{k}}\}$ that in the chiral $J=1$
state, due to the $p_{x}-ip_{y}$ structure of the order parameter,
in the presence of a vortex in the exciton condensate, guarantees
the existence of topologically protected midgap states bounded to
the vortex with energy $\Delta_{\mathbf{k}}^{0}$ \cite{read2000,gurarie2007,seradjeh2008}.

\begin{figure}[htb]
\includegraphics[width=8.11cm]{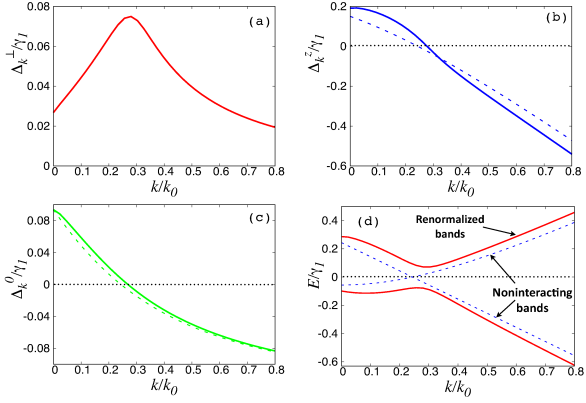}
\caption{(Color online). (a)-(c) $\left|\Delta_{\mathbf{k}}^{\perp}\right|,\Delta_{\mathbf{k}}^{z},\Delta_{\mathbf{k}}^{0}$
respectively, as a function of $k$ ($k_{0}\equiv\gamma_{1}/(\hbar v_{F})$),
for $T=0$, $d=1\text{nm}$, $\alpha=1$, and $V_{g}=0.3\gamma_{1}$.
In (b) and (c) the dashed lines show $\Delta_{\mathbf{k}}^{z}$
and $\Delta_{\mathbf{k}}^{0}$ respectively in the noninteracting
case. (d), The solid (dashed) lines show the renormalized (noninteracting) bands. 
\label{fig:deltas} }
\end{figure}

\begin{figure}[htb]
\includegraphics[width=8.11cm]{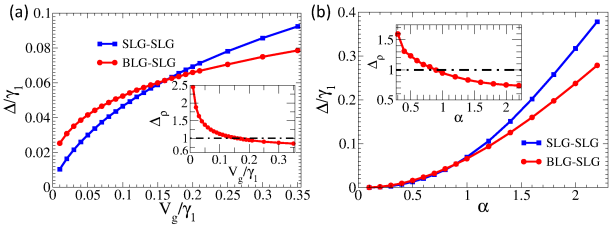}
\caption{(Color online). 
 $\Delta\equiv\left|\Delta_{\mathbf{k}}^{\perp}\right|_{\text{max}}$ as a function of 
 $V_{g}$, (a), and $\alpha$, (b), in the hybrid
 SLG-BLG structure and the symmetric SLG-SLG structure for $T=0$ and
 $d=1$~nm. In (a) $\alpha=1$, in (b) $V_{g}=0.2\gamma_{1}$.
 The insets show the ratio ($\Delta_\rho$) between $\Delta$ in SLG-BLG and $\Delta$ in SLG-SLG.
\label{fig:deltaVg} }
\end{figure}

From Fig.~\ref{fig:deltas} we see that, at $T=0$,
for typical parameter values the peak value
$\Delta\equiv\left|\Delta_{\mathbf{k}}^{\perp}\right|_{\text{max}}$
of the order parameter magnitude is $\approx0.075\gamma_{1}=30$~meV.
Fig.~\ref{fig:deltaVg}~(a) shows the dependence of $\Delta$
on $V_{g}$ for both SLG-BLG and SLG-SLG at $d=1\,\text{nm}$ and
$\alpha=1$, where $\alpha\equiv e^{2}/(\epsilon\hbar v_{F})$. 
We find that at low bias ($V_{g}/d<60\,\text{meV/nm}$) $\Delta$ is larger in the
hybrid SLG-BLG heterostructures.
Compared to the symmetric SLG-SLG
structure, in the SLG-BLG structure the density of states (DOS) in
one of the layers (BLG) is higher than in the SLG-SLG structure, and the
interlayer chiral factor $F^{d}(\theta_{\mathbf{k-p}})$ oscillates
more rapidly. The first effect favors the formation of the exciton
condensate and therefore enhances $\Delta$ whereas the second tends
to suppress it. We can then understand the scaling with $V_{g}$ of
the ratio ($\Delta_\rho$) between $\Delta$ for SLG-BLG and for SLG-SLG (inset of Fig.~\ref{fig:deltaVg}~(a))
as a result of the competition of two effects: the DOS effect
dominates at low $V_{g}$ and the fast oscillation of $F^{d}(\theta_{\mathbf{k-p}})$
takes over at high $V_{g}$. Fig.~\ref{fig:deltaVg}~(b)
also shows that
in the
weak coupling regime ($\alpha<1)$ the interlayer coherence can be stronger
in SLG-BLG than in SLG-SLG. 

The value of $\Delta$, for typical values of $V_{g}\approx0.3\gamma_{1}$,
suggests a {\em mean-field} critical temperature $T_{c}\lesssim300~K$.
This value is an overestimate. 
Because the system is two-dimensional
and the broken symmetry is $U(1)$, 
$T_{c}$ is reduced to the Berezinskii-Kosterlitz-Thouless
temperature ($T_{BKT}$) above which we have the proliferation of unbound
vortices and antivortices of the condensate. 
In addition, thermal and quantum phase fluctuations 
\cite{watchel2012},
screening 
\cite{kharitonov2008b,bistritzer2008,lozovik2009,kharitonov2010,lozovik2010,basu2010,basu2011,mink2012,lozovik2012,sodemann2012}
and disorder 
\cite{abergel2012c,bistritzer2008a}
can reduce considerably $T_{c}$. 
An accurate estimate of $T_c$ is beyond the reach of theory also due to
the uncertainties about the experimental conditions. However,
the degeneracy and chirality of the ground state are robust
and independent of the exact value of $T_c$. 
Screening and disorder are expected to be the dominant factors
in suppressing $T_c$  \cite{gorbachev2012}.
Screening in general will preserve the ``central'' nature of the 
interaction and therefore will not affect the degeneracy and chirality of the
phase coherent state. Similarly the presence of disorder will
renormalize the order parameter, and therefore $T_c$, but also do not
affect our main findings.
To show this, let us denote by a tilde the disorder-renormalized fields. For $\tilde\Delta_\kk^\perp$ we find:
\begin{equation}
\tilde{\Delta}_{\mathbf{k}}^{\perp}=\Delta_{\mathbf{k}}^{\perp}-\frac{n_{i}}{A}\sum_{\mathbf{p}}
 \frac{F^{d}(\theta_{\mathbf{k-p}}) U_{1}(\mathbf{k}-\mathbf{p})U_{2}^{*}(\mathbf{k}-\mathbf{p})
 \tilde{\Delta}_{\mathbf{p}}^{\perp}}{-\left(i\omega_{n}-\tilde{\Delta}_{\mathbf{p}}^{0}\right)^{2}+\left(\tilde{\Delta}_{\mathbf{p}}^{z}\right)^{2}+\left|\tilde{\Delta}_{\mathbf{p}}^{\perp}\right|^{2}},\label{eq:order_parameter}
\end{equation}
where $n_i$ is the impurity density, $U_\sigma$ is the disorder potential in layer $\sigma$,
and 
$\omega_n$ are the  Matsubara frequencies.
Eq.~(\ref{eq:order_parameter}) shows that the chiral factor $F^d$ appears in the same
way as in the gap-equation valid in the clean limit. This guarantees that
{\em even in the presence of disorder the chiral and the non-chiral sulutions are degenerate}
considering that for almost all cases of interest 
$U_\sigma(\theta_{\mathbf{k-p}})=U_\sigma(-\theta_{\mathbf{k-p}})$.

Considering that we find that in
SLG-BLG the mean-field $T_{c}$ for unscreened Coulomb interaction
is of the same order as in SLG-SLG and that screening, disorder, thermal and quantum fluctuations
are expected to affect $T_{c}$ similarly in the two systems, we conclude
that in realistic setups $T_{c}$ for SLG-BLG should be of the same
order as for SLG-SLG. Recent results \cite{gorbachev2012} show hints
of an exciton condensate for SLG-SLG in current experimental conditions.
We can then conclude that the combined effects of screening and disorder
in SLG-SLG and SLG-BLG heterostructures might suppress $T_{c}$ but
should not prevent the experimental observation of the predicted interlayer
phase coherent states.

In conclusion, we have shown that in hybrid heterostructures, in which
the electrons in different layers have different chirality ($m$ in
one layer and $n$ in the other) the interlayer phase coherent state
is 4-fold degenerate for short-range interactions, and 2-fold degenerate
for long-range ``central'' interactions when $m=2n$. Moreover,
we find that one of the degenerate states is always a chiral superfluid
state, a fact that implies the presence of protected midgap states
in the presence of vortices in the exciton condensate.
We also find that these properties of the ground state are robust
and are not affected by effects like screening and disorder that on
the other hand can strongly suppress
$T_{c}$ for the formation of the interlayer phase coherent state.

It is a pleasure to acknowledge Chris Triola for his support
in calculating the density-density response function, 
and Allan H. MacDonald and Shiwei Zhang
for very helpful discussions. Work supported by ONR, Grant
No. ONR-N00014-13-1-0321, and the Jeffress
Memorial Trust, Grant No. J-1033.
ER acknowledges the hospitality
of KITP, supported in part by NSF under Grant No. PHY11-25915, where
part of this work was done.





\end{document}